\documentstyle[multicol,psfig,aps,prl]{revtex}
\newcommand{\bmulticol}{\begin{multicols}{2}\narrowtext}
\newcommand{\emulticol}{\end{multicols}\widetext}
\begin{document}
\draft
\title{Noise-correlation-time-mediated localization in random
nonlinear dynamical
systems}
\author{Juan L. Cabrera $^{1,2}$, 
J. Gorro\~{n}ogoitia $^{1}$ and F. J. de la Rubia$^{1}$}
\address{$^{1}$Dpto. de F\'\i sica Fundamental, 
Universidad Nacional de Educaci\'on a Distancia,  
Apdo. 60141, 28080 Madrid, Spain\\
$^{2}$Centro de Astrobiolog\'{\i}a, INTA, 
Carretera de Ajalvir km.4, 28850 Torrej\'on de Ardoz, Madrid, Spain.
}
\date{To appear in {\it Phys.Rev.Lett.}}
\maketitle

\begin{abstract}
We investigate the behavior of the residence times
density function for
different nonlinear dynamical systems with limit cycle behavior and
perturbed parametrically with a colored noise.
We present evidence
that underlying the stochastic resonancelike behavior with the
noise correlation time, there is an effect of optimal localization of
the system trajectories in the phase space.
This phenomenon is observed in systems with
different nonlinearities, suggesting a degree of universality.
\end{abstract}

\bmulticol 
By stochastic resonance (SR) it is normally understood
the phenomenon by which an additive noise (usually considered
uncorrelated)
can enhance the coherent response of a periodically driven system.
First proposed in climate model studies \cite{Benzi}, SR was first
experimentally verified by Fauve and Heslot \cite{Fauve},
and since then this
behavior has been predicted and observed in many different
theoretical and experimental systems (see \cite{Gammaitoni1998}
for an extensive review and a complete list of references).
In particular, the presence of SR has been discussed in
a great number of models including
spatio-temporal systems \cite{Lindner1995}, and has helped to
understand how
biological organisms may use noise to enhance the transmission of weak
signals through nervous systems \cite{Douglass1993,Gluckman1996}.
Quite
recently it has been numerically shown that SR can also occur in
the
absence of an external periodic force as a consequence of the
intrinsic
dynamics of the nonlinear system \cite{Gang1993}, a behavior that has
been
denominated {\it autonomous stochastic resonance}.
Most of the work on SR has traditionally focused on
systems with additive
noise, and with some exceptions (see, for instance,
\cite{Gammaitoni1994})
little attention has been given to cases where the
noise perturbs the system parametrically, in spite of the well known
differences with the additive situation. With respect to non-white
noise,
the effect of additive colored noise on SR has been considered in
periodically driven overdamped systems \cite{Gammaitoni1989}, showing
that
the correlation time can suppress SR monotonically, a feature
demonstrated experimentally in \cite{Mantegna1995}. However, only
very recently the situation in which the system is subject to both
multiplicative and colored noise has been discussed in the literature.
In \cite{Fulinski} the authors analyze the effect of multiplicative
colored noise on periodically driven linear systems, discussing the
appearance of SR by changing either the intensity or the correlation
time of
the noise. For nonlinear models, in \cite{Cabrera1997} we
considered a
system without periodic external force but with an intrinsic limit
cycle
behavior, which was parametrically perturbed by an Ornstein-Uhlenbeck
(OU)
noise, finding a nonmonotonic behavior of the coherence in the system
response when measured as a function of the noise correlation time,
while no
coherence enhancement was obtained when changing the noise intensity.
A
similar result has also been recently obtained analytically for an
overdamped linear system periodically driven and parametrically
perturbed by
an OU process \cite{Barzykin98}.

In this paper, we present numerical evidence which suggests that
underlying the SR-like behavior as a function of the noise correlation
time,
there is a localization effect of the system trajectories in the phase
space
for a particular value of the correlation time. This is obtained in
systems
with intrinsic limit cycle, perturbed parametrically by
an OU process and with different nonlinearities, which is also a clear
indication that the phenomenon is not a peculiarity of an specific
model.

We study three different $2D$ random systems.
The delayed regulation model, known from
population dynamics \cite{Maynard1968}
\begin{equation}
x_{t+1}=\lambda _{t}x_{t}(1-x_{t-1}),  \label{drm}
\end{equation}
the Sel'kov model for glycolysis \cite{Selkov1968}
\begin{eqnarray}
\dot{x} &=&-x+\lambda _{t}y+x^{2}y,  \nonumber \\
\dot{y} &=&b-\lambda _{t}y-x^{2}y,  \label{selkov}
\end{eqnarray}
and the Odell model also from population dynamics
\cite{Odell1980}
\begin{eqnarray}
\dot{x} &=&x[x(1-x)-y],  \nonumber \\
\dot{y} &=&y(x-\lambda _{t}).  \label{odell}
\end{eqnarray}

\begin{figure}[h]
\[
\psfig{figure=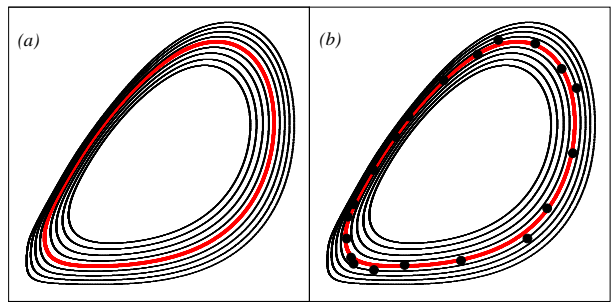,width=8cm}
\]
\caption[]{Phase space partition of Eq. (\ref{drm}), 
(a) with $N+1=9$ deterministic periodic 
attractors and (b) with superimposed random 
states (dots), and with $\sigma _{r}=0.05$ and $\tau =3$. 
The thick line corresponds to the attractor 
$\Gamma (\left\langle \lambda\right\rangle )$.}
\end{figure} 

\begin{figure}
\[
\psfig{figure=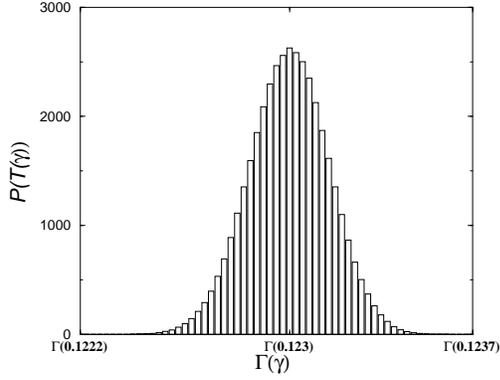,width=7cm}
\]
\caption[]{Residence times density function for 
(\ref{selkov}), obtained with $50$ realizations 
of $5\times 10^{7}$ time steps of
size $\Delta t=10^{-2}$, 
$\left\langle \lambda \right\rangle=0.123 $, 
$\sigma =5\times 10^{-4}$ and $\tau =6$.}
\end{figure} 

\begin{figure}
\[
\psfig{figure=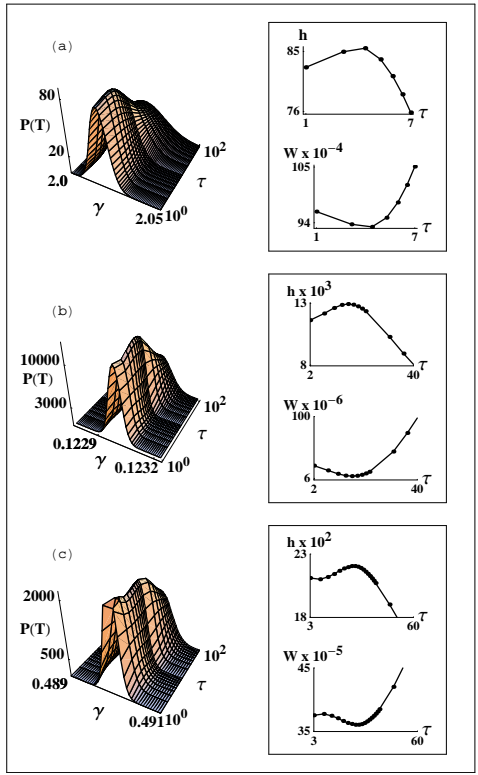,width=9cm}
\]
\caption[]{RTDF versus $\tau$ for (a) Eq. (\ref{drm}), 
(b) Eq. (\ref{selkov}), and (c) Eq. (\ref{odell}). 
Data obtained with (a) 
$\left\langle \lambda \right\rangle=2.02$, 
$\sigma _{r}=0.05$ and 100 realizations of $10^{6}$ iterations, (b)
$\left\langle \lambda \right\rangle =0.123$, $\sigma =5\times 10^{-4}$ 
and 50 realizations of $5\times 10^{7}$ time steps 
of size $\Delta t=10^{-2}$, and
(c) $\left\langle \lambda \right\rangle =0.49$, $\sigma =10^{-3}$ and
realizations as in (b). The inset plots indicate 
the height and width behavior.}
\end{figure} 
Here, $t$ takes discrete values in (\ref{drm}) or continuous values in
(\ref
{selkov}) and (\ref{odell}), and in all cases we will consider the
control parameter as a random variable
$\lambda_{t}=\lambda+\zeta _{t}$,
i.e., as a deterministic part $\lambda $, plus an stochastic
perturbation $\zeta _{t}$, which is assumed to be an OU
process, i.e., a stationary Gaussian Markov
noise with zero mean,
$\left\langle \zeta_{t}\right\rangle =0$,
and exponential correlation
$\left\langle \zeta_{t}\zeta _{t^{\prime }}\right\rangle =
\left( D/\tau \right) \exp\left(-\left| t-t^{\prime }\right| /\tau
\right) $,
where $\tau $ is the correlation time and $D/\tau =\sigma ^{2}$
is the variance of the noise.
We will refer to the square root of the variance,
$\sigma $,
as the intensity of the noise. The deterministic counterparts of
(\ref{drm}), (\ref{selkov}) and (\ref{odell}) undergo a
supercritical Hopf bifurcation
at $\lambda \equiv \lambda _{H}$ which, in the Sel'kov
model, also depends on the parameter $b$.

The numerical integration has been carried out
with $\lambda $ in the
limit cycle parameter domain.
The iteration of (\ref{drm}) has been
recreated using an integral algorithm \cite{Fox1988} that guarantees
the
quality of the correlation function in the simulations of the noise at
discrete times, while (\ref{selkov}) and (\ref{odell}) have been
integrated
by an order $2$ explicit weak scheme \cite{kloeden}. The results
presented
hereafter are independent of the initial conditions and were obtained
after
the decay of the initial transients.

The observed fact \cite{Cabrera1997} that for a particular correlation
time $\tau _{r}$ the coherence of the
system oscillations has a maximum, and that
the frequency of these oscillations is close to the
deterministic one $w_{d}$, seems to indicate that
for the resonant correlation time, the
probability that the system visits the attractors associated with the
mean control parameter value $\lambda =<\lambda _{t}>$ has also a
maximum.
If this is the case, this maximum should be accompanied with a decrease
in
the probability to visit other attractors associated with parameters
far away from $\lambda $, or, in other words, should lead to an effect
of
{\it concentration} or {\it localization} of orbits around the
attractor
associated with $\lambda $ as soon as $\tau \sim \tau _{r}$.
It is worth recalling that because of changes in the stability
properties, a
somehow similar localization effect can also occur in parametric
deterministic systems with time dependent parameters, as is the case,
for instance, in the well known parametric resonance phenomenon.

With the aim of studying the residence times distribution of the system
on the different available attractors in the system periodic or
quasiperiodic domain, we consider a deterministic attractor $\Lambda
(\lambda )$, i.e., the attractor obtained with the deterministic
counterpart
of the stochastic system, evaluated at a particular value of the
control
parameter, $\lambda $. Next we divide the system phase space in $N+1$
attractors associated with $N+1$ values of the parameter separated a
distance $\Delta \lambda $. In this way, a mesh is composed by
concentric
deterministic attractors centered around the stationary equilibrium
state 
$(x^{*},y^{*})\mid _{\lambda \sim \lambda _{H}}$, with $\lambda$
in the fixed point domain. This partition looks like the one shown in Fig.
1a. With
this construction, we have a series of $N+1$ attractors $\{\Lambda
(\lambda
_{N/2_{-}})...\Lambda (\lambda _{1_{-}}),\Lambda (\lambda
_{0}),\Lambda
(\lambda _{1_{+}})...\Lambda (\lambda _{N/2_{+}})\}$, where we use the
definition $\lambda _{k\pm }\equiv \lambda \pm k\Delta \lambda $. This
series divides the phase space in $N$ rings, each one denoted by
$\Gamma
(\gamma _{k})\equiv (\Lambda (\lambda _{k}),\Lambda (\lambda
_{k+1}))$,
where $\gamma _{k}\equiv \left( \lambda _{k+1}+\lambda _{k}\right) /2$
is
the mean control parameter obtained with the control parameters that
define
the boundary of the ring. The stochastic system is integrated on this
mesh,
and its evolution describes random trajectories as the one described
in Fig.
1b for the particular case of (\ref{drm}), visiting during a finite
time
each ring of the mesh. During the integration process we measure the
residence time in the rings as follows: let $t_{1}^{k}$ and
$t_{2}^{k}$ be
the entrance and exit times to the ring $\Gamma (\gamma _{k})$,
respectively. The residence time in this ring is $t(\gamma
_{k})=t_{2}^{k}-t_{1}^{k}$, and we denote the residence time of the
$n$
visit event to the ring $\Gamma (\gamma _{k})$ by $t_{n}(\gamma
_{k})$.
Then, if during an integration time $I$, which is achieved by
integrating $R$
realizations of $M$ time steps, there have been $V_{k}$ visit events
to the
ring $\Gamma (\gamma _{k})$, the mean residence time of the system in
this
ring is given by the mean of the residence events, that is, $T(\Gamma
(\gamma _{k}))\equiv \sum_{n=1}^{V_{k}}\frac{t_{n}(\gamma _{k})}{I}.$
Such a
determination of the residence times gives an alternative statistical
measure of the resonant amplification described in \cite{Cabrera1997}.
Therefore, given a pair $(\sigma ,\tau )$, the function defined by the
histogram
$P(T)\equiv P(T(\gamma _{k}))\equiv P(\frac{T(\Gamma
(\gamma_{k}))}{\Delta \lambda })$
is a measure of the probability density for the
system
state to be in the region defined by the ring $\Gamma (\gamma _{k})$ .
An
example of histogram is depicted in Fig. 2 and shows that the system
mostly visits the attractors surrounding the ring $\Gamma (<\lambda
_{t}>)$.
We remark that we have carefully selected the simulation
parameters to ensure that the partition does not
contain
overlapped attractors such that this has a well defined meaning. 
An illustrative example of the residence times
density function (RTDF) as a function of
the correlation time is depicted in Fig. 3 for the three models.
Obviously,
the localization of the system trajectories depends strongly on $\tau$.
The
RTDF height shows a nonmonotonous behavior reaching a maximum at a
particular value of $\tau \sim \tau ^{*}$ and, at the same value, the
width $W$ calculated at the height $h/\sqrt{e}$ shows a remarkable
minimum,
as
represented in the inset curves. {\it The correlation time of the
parametric
random perturbation acts as a tuner which controls (in an statistical
sense)
the behavior of the system, maximizing its localization on the region
of the
phase space surrounding $<\lambda _{t}>$}. Furthermore,
the relation
$h/W$ has a maximum for a particular value of $\tau $, and this optimal
value
depends on $\lambda =<\lambda _{t}>$,  as can be appreciated in Fig.4.
Such a dependence enables us to relate the optimal correlation time
for
maximal localization, $\tau^{*}$, with the temporal scales of the
deterministic counterparts. We  first study the behavior of the
postponement of the bifurcation point because of the multiplicative
noise in
order to obtain the postponed bifurcation point
$\lambda_{H}^{*}(\sigma,\tau )$.We next calculate
the effective distance to the bifurcation
point $\Delta \lambda ^{*}=|\lambda -\lambda _{H}^{*}|$, and measure
from the
deterministic temporal series the period, $T^{*}$, of the oscillations
when
the system is evaluated at a distance $\Delta \lambda ^{*}$ from the
deterministic bifurcation point. With this information, in Fig. 5 we
plot
the behavior of $\tau ^{*}$ with the quantity $\Delta T^{*}\equiv
|T^{*}-T(\lambda _{H})|$, where $T(\lambda _{H})$ is the period of the
deterministic system at precisely the Hopf bifurcation point. The
curves can
be fitted by a power law $\tau^{*} \sim \left( \Delta T^{*}\right)
^{\alpha }$
with the exponents $\alpha =-0.59$, $-0.58$ and $-0.53$ for
(\ref{drm}), (%
\ref{selkov}) and (\ref{odell}) respectively, and this seems to
indicate
that the localization behavior with $\tau $ is characterized by a
unique
exponent with value close to $-1/2$. We note that for the case of
(\ref{drm}%
) it is even possible to relate $\Delta T^{*}$ with the system
implicit
periodicity $T_{2}$, thus recovering a similar relation to that
calculated
in \cite{Cabrera1997}. In this way, these results relate the
resonantlike behavior previously reported in \cite{Cabrera1997}
using the quality factor $\beta$ \cite{quality},  
with an increase (in mean) of the localization of
the orbits of the system. 
\vspace{1cm}
\begin{figure}
\[
\psfig{figure=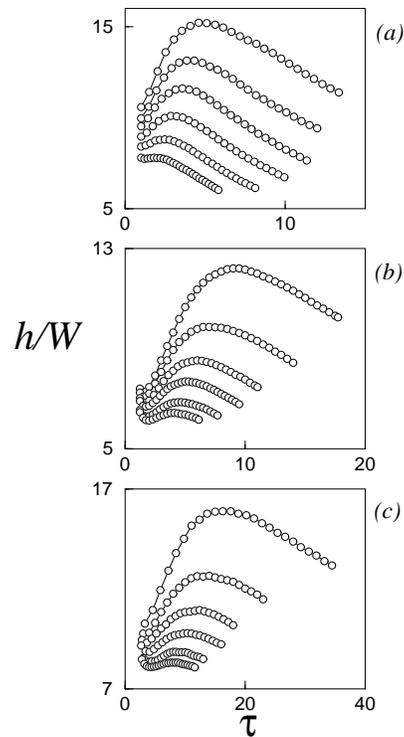,width=12cm}
\]
\caption[]{$h/W$ versus $\tau $ for
different $\left\langle \lambda \right\rangle $ for (a) Eq. (\ref{drm}) 
($h/W \times 10^{3}$, $\sigma_r=0.05$ and, 
from top to bottom $\left\langle \lambda \right\rangle = 
2.00828, 2.01055, 2.01344, 2.01713, 2.02182$ and $2.02781$), 
(b) Eq. (\ref{selkov}) ($h/W \times 10^{6}$, $\sigma=5 \times 10^{-4}$ 
and, $\left\langle \lambda \right\rangle = 0.124, 
0.1235, 0.123, 0.1225, 0.122$ and $0.1215$), and (c) Eq. (\ref{odell})
($h/W \times 10^{5}$, $\sigma=10^{-3}$ and, 
$\left\langle \lambda \right\rangle = 0.496, 
0.494, 0.492, 0.49, 0.488$ and $0.486$). Number of realizations, 
iterations and time steps are as in Fig. 3.}
\end{figure} 
From the behavior of the quantity $h/W$, 
it is clear that a concentration of orbits
around a narrow range of bands in the phase space implies a bigger
weight of
those particular frequencies in the power spectrum, and, as a
consequence, a
nonmonotonous behavior qualitatively similar to that of Fig. 4 should
be expected for $\beta$, indicating an increase of the coherence
in the
system response. This is indeed the case for our three models
(with quality factors showing a maximum
for values of the correlation time close to
$\tau^{*}$) clearly indicating that the SR-like effect
induced by colored noise in nonlinear
systems with limit cycle behavior is quite general.

Summarizing, we have presented
numerical evidence of a novel
effect of enhanced localization of orbits
mediated by the correlation
time of a multiplicative OU process in nonlinear
dynamical systems with limit cycle behavior.
This effect is characterized by a power
law with exponent close to $-1/2$
for all the models considered in spite of
their different nonlinearities.
This behavior could indicate the universal
character of this phenomenon, but
further research is required to clarify
this point. This work also relates the SR-like behavior
previously reported with this localization effect. 

We acknowledge financial support from DGESEIC (Spain) project
PB97-\-0076.
\vspace{1cm}
\begin{figure}
\[
\psfig{figure=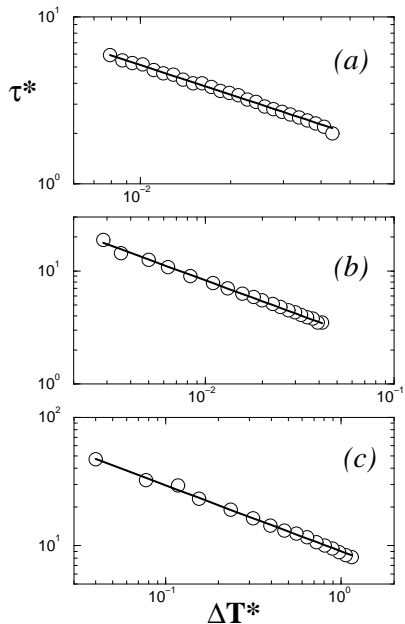,width=10cm}
\]
\caption[]{$\tau ^{*}$ versus  $\Delta T^{*}$ for 
(a) Eq. (\ref{drm}), (b) Eq. (\ref{selkov}), and (c) Eq. (\ref{odell}).
Simulation parameters are as in Fig. 3.}
\end{figure}

\emulticol 

\end{document}